\documentclass[12pt,english]{article}
\usepackage{graphicx}
\usepackage[english]{babel}
\usepackage[ansinew]{inputenc}
\usepackage[T1]{fontenc}
\usepackage{latexsym}
\usepackage{amssymb}

\begin{document}
\title{
Liquid scintillator as tracking detector for high-energy events}
\author{Juha Peltoniemi\\ \em Neutrinica Ltd, Oulu, Finland\\ and 
\em Excellence Cluster Universe \\ \em Technische Universität München, Garching, Germany}
\date{\today}
\maketitle

\begin{abstract}

A large-volume liquid scintillator can be used as a tracking detector to measure high-energy neutrino events, like atmospheric neutrinos and neutrino beams. The lepton flavor recognition is almost absolute above 1 GeV. The energy resolution is  2--5 \%, the main uncertainties coming from nuclear physics and poorly recognized hadrons. At GeV scale antineutrinos may be statistically distinguishable from neutrinos by neutron and proton signals. 
\end{abstract}

\section{Introduction}

This work is motivated by LENA experiment which has been proposed to study proton decay and neutrino astronomy \cite{Wurm:2007cy, Oberauer:2006cd, Hochmuth:2006gz, MarrodanUndagoitia:2006qs, MarrodanUndagoitia:2006qn, MarrodanUndagoitia:2006rf, MarrodanUndagoitia:2006re, Undagoitia:2005uu, Autiero:2007zj}. 
Although LENA is designed for low-energy neutrinos, observing high-energy neutrinos would increase its physics range. So far the performance at GeV range or above has not been simulated exactly. Some studies on through-going muons have been made for Borexino \cite{wurm} and some others, and parallel studies \cite{Learned:2009rv} on high-energy neutrinos are being made for HanoHano in Hawaii \cite{Batygov:2008mr}. 

The liquid scintillator is prominent for its capacity to measure energy. In low energies the typical energy resolution of liquid scintillator (like Borexino) is 
\begin{equation}
\frac{\delta E}{\rm MeV} = 5 \% \sqrt{\frac{E}{\rm MeV}}
\end{equation}
which is what LENA is targeted to be about. At high-energies there will be other effects that spoil this. A particular effect is caused by quenching, because of which particles with different momenta and different masses scintillate differently. Hence it is very important to count and identify all particles.

There is also a major difference between small and large detectors, namely when the detector dimension is comparable to attenuation length, the total scintillation light may differ more than a factor of 2 depending on the location of the event within the detector, and an overall 1 \% energy resolution may require 10--20 cm positional resolution near the borders.
  
The purpose of this study is to quantify those phenomena affecting the energy resolution in LENA.

The high-energy performance is relevant for atmospheric neutrinos and neutrino beams. The beams may include
wide band beam, beta beam and  neutrino factory.
The last one requires lepton charge recognition. Here we consider the possibility to deduct the charge by recognizing recoil nucleons. Magnetic detectors are not discussed in this note.

For the baseline CERN-Pyh\"asalmi of 2288 km \cite{Peltoniemi:2006hf} the optimal energy is around 4.2 GeV. This baseline allows to define the mass hierachy and study CP violation, depending on neutrino parameters and performance of the experiment. To fully exploit the beam, good performance at 1--5 GeV range is required. For shorter baselines we may need sub-GeV performance. However, baselines shorter than 1000 km lack capacity for mass hierarchy and consequently CP violation.

The interesting and observable atmospheric energy spectrum extends from tens of MeV to tens of GeV. Sub-GeV neutrinos may provide more detailed information on neutrino oscillation, but an accurate measurement of multi-GeV neutrinos is also important for precision measurement of oscillation parameters and defining the mass hierarchy in case of favourable parameters. There is a particular window of low-energy muons that cannot be covered by Water Cherenkov detectors, and is hard with iron calorimeters. 

Higher energy neutrinos may also be a signal of something more exotic, like an indirect evidence of dark matter (annihilation in the Sun). Neutrinos of cosmic origin are beyond reach and interest for this kind of experiments. 

The purpose of this study is to find out whether it is possible to reconstruct the original neutrino event and the tracks associated to it. It is also expected to give credible hints on the performance, particularly on energy resolution. While more exact results may require years' work on simulation, measurements and calibration, the present status of these studies is presented, as it may have implications on strategic decisions on future experiments and their design.

\section{Physics}

The scope of this works extends to neutrino energies 1--5 GeV. Some extrapolations will be done below and beyond this range.

The event types with the relevant energies include quasielastic scattering ($
\nu_\mu +n \to p + \mu^-$ and $\overline{\nu}_\mu +p \to n + \mu^+ $),
resonant single pion production ($\nu_\mu + n \to  \Delta^+ + \mu^- \to  p+ \pi^0  + \mu^- $ etc.)
and deep inelastic scattering, at these energies essentially a multipion event (e.g. $\nu_\mu +n \to  p + N_\pi \pi + \mu^-$). DIS and SPP may have rather similar signatures.  All processes (QES, SPP, DIS) occur at least at 10 \% fraction throughout the energy range (1--5 GeV). QES dominates at 1 GeV and below, SPP around 2 GeV and DIS above 3 GeV.

With a bound nucleon all kinds of nuclear fractions may occur in final states, as well as unnumerable amount of gammas, typically in MeV scales. Of the nuclear residues, loose neutrons will give identifiable signals, but low-energy protons and alphas may not be distinquished at all, although they contribute to scintillation light. The binding energy and other unmeasurable losses in nuclei give rise to an unavoidable systematic uncertainty in the result, as there is little chance to deduce whether the scattering occurred from carbon or hydrogen. These effects are absent below 300--500 MeV where the energies are insufficient to expel a nucleon from the nucleus, and the main uncertainty comes from the unmeasurable momentum of the recoil nucleus.

Neutral pions decay immediately ($8\cdot 10^{-17}$ s) into two photons ($\pi^0 \to \gamma\gamma$). Photons are invisible until they interact with pair production, with a mean free path of about half a metre, resulting in an electromagnetic shower equivalent to a double electron shower. The charged pions decay ($\pi^+ \to \mu^+ \nu_\mu$) typically at rest, with mean lifetime 26 ns, but the signal of the slow muon (4 MeV) is hardly separable from the afterglow of the main event. The muon decay ($\mu^+ \to e^+ \nu_e \overline{\nu}_\mu $) instead, with lifetime 2 $\mu$s (at rest), is well separated from the original event, and easily recognizable and localizable unless the electron remains at rest. The muon decay can hence be used to identify the particles ($\mu$ or $\pi$) and define the endpoint of the track.

Neutron captures should be visible. (Neutrons have not yet been studied explicitely).
Observing and localizing the capture of the prompt neutron will help to define the event topology and to measure the energy. Identifying the prompt nucleon as proton or neutron can be used to define lepton charge without magnetic field, but this is quite unreliable at higher energies, as already in SPP process either nucleon may be in the final state, and high energy interaction can also expel additional neutrons from the nucleus.

\section{Simulations}
\subsection{Codes}

For the first approximation the detector performance was simulated with a home-made Java code "Scinderella"
\cite{scinderella}. An improved integrated codebase is under development.

The detector model used in the simulations is a simplified version of LENA, ignoring the top and bottom photosensors. 
The total photon efficiency with used assumptions is about 1--2 \%, depending on the location relative to walls (Borexino is 3 \%). The attenuation length is assumed to be 15 m and time jitter/resolution of photosensors 1 ns, and the scintillation decay properties as in \cite{MarrodanUndagoitia:2009kq}.

The events are so far generated using internal event generator.
The light emission is modelled analytically, with photon directions and absorption in medium considered stochastically. 

The analysis are done using both least squares and maximum likelihood fit. The latter is more robust and converges much better.

\subsection{Limitations}

These studies have currently the following physical limitations, to be included in subsequent work:
\begin{itemize}
	\item DIS events are not yet very realistic and physical.
	\item Hadron interactions are not taken into account. At these energies the DIS event is not really a shower, but just a multipion event, neutral pions decaying to two electromagnetic showers. The typical pions tracks are comparable to the hadron interaction length 92 cm, but the pion energies of a few tens or hundreds of MeV's are not sufficient to generate further pions. Elastic or semi-elastic collisions may, however, lose energy to recoils or nucleons kicked-off from nuclei. Such collision may spoil the analysis but might also help to identify particles. Pions may also get absorbed, the probability of which is charge dependent.
	\item Neutrons are not followed. For high energy collisions the neutron tracks may be several metres, and evidently followable (kicked protons scintillate, though quenched). Also the absorption signal, a few MeV gamma with a mean delay of 200 $\mu$s, can be seen within the scintillator or in the buffer, unless the neutron escapes the detector. A high-energy neutron generates additional secondary neutrons by spallation reactions.
	\item Fluctuations on showers are not correctly modelled. Also non-showering low-energy electrons ($E<200$ MeV) and gammas are not modelled correctly. Sub-GeV are already doubtful.
 	\item Transversal evolution of electromagnetic shower is not taken into account. The width is typically 9 cm, hidden under the time resolution of PMT's. With good statistics or better trackability, however, it could be partially measurable and used for identification, if necessary. 
	\item Multiple Coulombian scattering of muons is neither taken into account --- it will give some stochastic transversal displacements, relevant for muons in magnetic detector only.
	\item Scattering of photons from the scintillator, other photosensors or back walls is not taken into account. That may dilute and lengthen the signal. In principle it is doable. 
\end{itemize}
Also, the saturation, deadtime, non-linear response of detector components are not included. This might be dangerous, and avoiding these puts stringent requirements for the design. The afterpulses and other irregularities within photosensor are neither taken into account, but they are probably manageable.

\section{Results of simulations}

\subsection{General notes}

For 1--5 GeV neutrino events there will be hundreds of thousands of photons. Hence the measurement is certainly not limited by statistics. At sub-GeV the statistics may limit the analysis, though the shortness of tracks poses even harder a challenge.

Some properties of the detector have great influence on the accuracy. Particularly important is the time behavior of the scintillator. The fast decay time is indeed much more important than light output or attenuation length.

The event topology was well identified in the simplest cases. Single pion events are no problem, and protons were usually recognized correctly. Multi-pion events impose major challenges, and it is hard to get all the tracks. Particularly in electron neutrino events most of the lepton energy is deposited near the vertex, and it is difficult to distinguish low-energy gammas and protons from the energy mess. 

The easiest topologies are those with large angles. A multiparticle decay at rest is the very most trivial case.

When using the known event as the initial value for the test event, the true solution was found to be very robust and stable. The error bars for energy were quite small.  

Neutrino energies higher than 5 GeV have not been considered explicitely. From these experiences we may expect that:
\begin{itemize}
	\item The hadron showers produce a mess of particles, from which one cannot distinguish individual tracks.
	\item The total energy is always definable with a few per cent uncertainty.
	\item The neutrino direction can also be defined rather accurately.
	\item Most eventa are partially contained but they may be partly studiable, as the energy of secondary particles correlates with lepton energy.
\end{itemize}

\subsection{Particle identification}

The lepton flavor recognition above 1 GeV is almost absolute, even without using decay signatures. Only some very rare events may fake wrong flavors ($\pi$ for $\mu$ or $\gamma$ for $e$), and most of such events are rejectable as they are a mess anyway, giving a poor fit. With a wide band beam of 1 \% intrinsic beam background the misidentifications are ignorable, but they might have some minor relevance for beta beams, even more for neutrino factory.

Charged pions (except charge) can be recognized by the decay signature and path length. In rare cases they may be confused with muons, because of similar decay signature, but usually the muon path is considerably longer. Single pion production by neutral current background is, however, a different case. In practice that is passed as background to further simulations.

High-energy protons are distinguishable by the characteristic rise of energy deposit at the end, even though quenched. As the slowest protons lose energy most rapidly, the low-energy  proton tracks are very short, remaining indistinguishable (below 100 MeV). The same is true for all nuclear recoils and fractions like alphas, for which only the energy can be measured, quenching giving major error, but direction remains completely unknown. 

The absorption signal is used to recognize neutrons. The observation probability can be as high as 95 \%. Neutron passage will also give a characteristic set of signals, but due to stochasticity it is very hard to make any adequate fit.

The neutron and proton signals can be used to distinguish neutrinos from antineutrinos, though only for events without charged pions and additional nuclear spillouts. The signature is completely lost if there are charged pions, this case being separable by decay signatures. The fraction of useful events at 1 GeV is max 80 \%, at 3 GeV 40 \%, at 5 GeV 20 \%, decreasing to below 10 \% at 10 GeV. The nucleon recognition for good events can be taken to be about 95 \%.

For low-energy scattering without nucleon emission the beta decay signature of $\mbox{}^{12}$B may be used to identify antineutrinos. Other beta active nuclei decay with electron capture which does not give detectable signal.

The direct charge identification is beyond the scope of this work.   

\subsection{Spatial and angular resolution}

The correct position of the vertex was located at a few-cm scale, if the correct minimum was found.  The location of the vertex remains sufficiently accurate for all the studied range. 

The angle of a single muon or electron above 1 GeV can be measured at a precision better than about 1 degree (0.015 rad), depending on case. The angular resolution is gradually weakened below 1 GeV, but still 100 MeV electrons fit much better than 0.1 rad (6 degrees). Also at high energies, despite all the mess, the muon direction can be measured better than 0.01 rad.

The angular resolution for other particles is substantially poorer: 
\begin{itemize}
	\item Higher-energy pions can be fitted at 0.1 rad accuracy.
	\item High-energy protons can be fitted at similar accuracy, but low-energy protons (less than 100 MeV, with sub-10 cm track) cannot be fitted at all. Below incident energies of 1 GeV the recoil momenta are practically unmeasurable.
	\item Gammas (gamma showers) in multiparticle events are poorly measurable, because they are easily hidden under other tracks. Only in cases of well-separated high-energy gammas from one decaying $\pi^0$(or in absense of other particles) can they be identified and measured, at 0.1 rad accuracy.
\end{itemize}
The accuracy of the definition of the direction of the incoming neutrino is basically limited by these recoil and secondary particles. This is very event-dependent, the angular resolution being typically better than 0.1 rad above 3 GeV and worse than 0.2 rad below 1 GeV. 

For highest-energy DIS events it is impossible to fit all tracks. However, the total momentum with an approximated fit allows to define the incident neutrino direction better than 0.1 rad accuracy, sufficient for atmospheric neutrino studies. 

\subsection{Energy resolution}

The energy resolution in its brutest form is given by the light sum. The location of the track is important as the light collection may vary a factor 2 depending whether the event is in the middle or in the border of the fiducial volume. The total deposited energy is defined better than the energy of the lepton, in most cases better than 1 \%. With a set of runs, the mean deviation with good fits was about 0.5 \%. 
	
For energetic DIS showers the energies of individual particles (all others than the lepton) are hard or impossible to measure. The total energy can be measured more accurately, however, even if the contents remain unidentified. The fluctuations for individual tracks are very large but they are reduced statistically.
	
The precision to define the energy of the incoming neutrino is hence mostly limited by the recognition of individual final state particles, as well as other physics not properly accounted for, like nuclear effects and instrumental effects in the detector. Compared with these, the total deposited energy can be considered almost absolute.

For beam coming from a fixed direction we can use the measurement of the lepton direction to define the neutrino energy. For quasielastic scattering, the energy of the charged lepton can be given as
\begin{equation}\label{ares}
 E_\nu = \frac{2(M-E_B)E_\mu - E_B^2+2ME_B - m_\mu^2 + \Delta M^2}
                    {2\left[(M-E_B)-E_\mu + p_\mu \cos\theta_\mu\right]}
\end{equation}
where $E_B$ is the nuclear binding energy (16--37 MeV for carbon, 0 MeV for protons), $M$ is the target mass, $\delta M^2$ the difference of square masses of the target nucleon and final state nucleon, $\theta_\mu$ is the lepton scattering angle. In principle this can be used for single pion production using $\Delta$ as the final nucleon, but with a lot of care due to other complicacies. For DIS scattering this makes no sence.

Typical accuracy by energy measurement using (\ref{ares}) is a few per cent. In practice this relation is relevant only for quasielastic scattering with poorly measured neutron or low-energy proton. Otherwise the measured deposited energy gives a better estimate for the neutrino energy.

For all cases there remains the uncertainty related to nuclear physics. The binding energy of carbon is 16.0 MeV and 18.7 MeV for the least bound proton and neutron, and 37 MeV for the s-state (leaving the nucleus in excited state). In Fermi bag model, the Fermi energy of carbon is 38 MeV, and Fermi momentum $221\pm 5$  MeV. Further nuclear excitations and breakups may absorb some hidden energy. We have little chances to define whether the scattering occurred from carbon or hydrogen, some additional energy associable to the verted being the only hint. 
Hence there is always a term ca 40 MeV in the energy resolution, which may be reduced no more than half if we can measure some released nuclear energy.

\section{Implications of the results of simulations}

\subsection{Neutrino beams}

The main purpose of this study is to find the performance numbers to be used in neutrino oscillation simulations, e.g. with GLoBES\cite{Huber:2007ji,Huber:2004ka}. 
This is far from trivial, and not all input required by GLOBES can be defined quantitatively.

In the truest case one should treat all events idividually, assigning energy resolution event-by-event base. While this is possible and mandatory in the analysis of the measured data, it is unpractical at this stage. One resolution function is too simple, however, so the events should be divided in at least three categories: the good (clean quasi-elastic and single-pion events), the bad (unresolvable mess) and the ugly (recognizable multiparticle showers). This kind of separation is important because for
the GLOBES simulations --- as well as for any data analysis --- the events with best accuracy count much more than their fraction, but we want to extract all possible information also from the very worst events.  

Assuming optimistically but realistically that the hardware will be good enough and the further simulations show no negative surprises, I suggest to use so far for the energy resolution at 1--6 GeV the following contributions:
\begin{enumerate}
	\item Statistical and systematical uncertainties for the light output and track reconstruction analysis:
	\begin{eqnarray}
\delta E &=& 0.01 E 
\end{eqnarray}
\item Unseen or unrecognized nuclear recoil and low-energy particles including gammas and secondary collisions:
\begin{eqnarray}
\delta E_{e} &=& 0.03  \mbox{ GeV}\sqrt{\frac{E}{1 \mbox{ GeV}}}\\
\delta E_{\mu} &=& 0.01  \mbox{ GeV}\sqrt{\frac{E}{1 \mbox{ GeV}}}
\end{eqnarray}
This is larger for electrons where energy deposit is closer to vertex, and fluctuations may hide other particles.
\item Unmeasurable prompt neutron energy (if neutron)
\begin{equation}
\delta E_{n} = 0.05  \mbox{ GeV}\sqrt{\frac{E}{1 \mbox{ GeV}}}
\end{equation}
This might be reduced with a better neutron tracking procedure.
\item Nuclear physics
 \begin{equation}
\delta E_{\rm nuc} = 20 \mbox{ GeV}.
\end{equation}
Assuming that most de-excitation gammas produce visible energy. This term is absent below O(500 MeV) where nucleons are not kicked out of nuclei.
\end{enumerate}
Optimist may halve the total resolution, pessimist double.

The background due to misidentification of lepton flavor can be taken zero, at least for a conventional beam.

The charge identification by neutron capture signal can be used to reduce the wrong sign muon background for muon disappearance studies in conventional beams. The miss probability is 5 \% for the good events (no charged pions and other messes) but should be taken full 50 \% for ugly events. For the neutrino factory this is far from required. 

Neutral current pion production gives small but non-ignorable background for muon flux. This depends on the beam spectrum and should be defined separately.

\subsection{Atmospheric neutrinos}

The case of atmospheric neutrinos differs from beam neutrinos as the direction of the incoming neutrino is also to be determined. As the azimuthal angle defines the path length $L$ through the Earth, it is equally important as energy for defining the oscillation phase proportional to $L/E$. 

As the energy of the first maximum is given by
\begin{equation}
E = (1.8 \pm 0.1) \mbox{ GeV} \left(\frac{L}{1000 \mbox{ km}}\right)
\end{equation}
the relevant energies are up to 23 GeV for upward going neutrinos, 1 GeV for horizontal neutrinos and 40 MeV for downward neutrinos.

For atmospheric neutrino studies we may thus distinguish three interesting ranges:
\begin{enumerate}
	\item 10--20 GeV. At these energies the lepton paths are long, up to 100 m for muons, hence only neutrinos parallel to detector can be studied. The vertical orientation is particularly suitable as upward going neutrinos are most interesting. The fiducial volume and angular aperture are greatly reduced.
	
	The energy resolution is bound by effetive recognition of the hadron shower. Nuclear effects play minor role. The muon track can be considered absolute, and the electron track is also very accurate. As the hadron shower is also very forward peaked, the angular resolution is good. The resolution in $L/E$ can be assumed to be better than 5 \%.
	
	The flavor recognition can be assumed absolute, but there is no capacity to define the lepton charge, even statistically. 
	\item Around 3 GeV: The flavour recognition is very good, and neutrinos can be distinguished from antineutrinos with a fair probability. Energy resolution is some 5 \%, nuclear effects being important. Angular resolution is not as good, worse than 0.1 rad, as there may be substantial undefinable momentum in short-track hadrons or nuclear fractions. 
	
		For upward going neutrinos, however, the requirement for directional accuracy is not too stringent, and the tracking accuracy may be sufficient. At least we can see clearly the 4th maxima and minima, maybe also 5th.	
	For horizontal and inclined angles the angular resolution for incident neutrino is not sufficient to determine $L$ at event-by-event basis. Hence $L/E$ analysis for these angles must be done statistically. 
	
	\item  Around 100 MeV. At these energies the solar oscillation parameteres contribute, and the oscillation pattern may be very interesting \cite{Peres:2009xe}. Charge recognition by neutron capture is lost when energy transfer is insufficient to bounce neutrons, but $\mbox{}^8$B beta decay may give a signal. The energy resolution is better than that at higher energies as nuclear effects do not contribute that much and there are no additional particles. Flavor recognition remains good, but not absolute. The angular resolution for lepton track is poor, and completely lost for nuclear recoils, so that the neutrino direction can be defined at hemispheric accuracy. As the scatterings can be rather isotropic, the $L/E$ analysis can be done only statistically. 
\end{enumerate}

The muon decay signal gives sufficiently reliable particle identification for all cases.

\section{Conclusions}

A large volume liquid scintillator turns out to be a viable candidate for a high-energy neutrino detector. It is a very economical choice because of its relatively low cost and ability to measure all kinds of neutrinos from different sources. If LENA will be built for neutrino astronomy, not using it for beam would be a waste of an opportunity, particularly because the baseline CERN-Pyh\"asalmi provides a good performance for observing long baseline observations.

The particle identification capacity and the energy resolution turn out to be very good. Throughout the considered range the accuracy for defining the energy of the incident neutrino is better than 5 \%, with clear events could be up to 2 \%. The lepton flavor is recognized almost always. For a limited set of events antineutrinos can be distinguished from neutrinos, though not sufficiently reliably for neutrino factories. 

For a wide band beam LENA as proposed is good enough. The vertical orientation is not too big a burden when the main signal is electrons with shower extending to a few metres. A vertical LENA serves also for beta beam, with a penalty of efficiency for highest energy neutrinos. Horizontal direction would be clearly better for muon neutrinos above 3 GeV, however. Some fiducial volume may be regained by using the external buffer for measuring the tails. It is also rather easy and cheap to design and build a dedicated auxiliary detector for 5 GeV muons.

LENA performs very well for atmospheric neutrinos, too. While the size is similar to Super-Kamiokande, LENA will have advantage because of better resolution and limited charge identification. The vertical orientation of the cylinder is optimal for measuring the first oscillation maximum. Moreover, liquid scintillator can measure muon neutrinos between 150--200 MeV, inaccessible for water Cherenkov detectors.

Performance in high energies poses challenging requirements to the detector. First of all, PMT's need good time resolution and non-saturating multi-photon capacity with not too many afterpulses or dead time. The scintillator liquid has to be as fast as possible, with little scattering. Light emission and attenuation length are secondary requirements. Evidently not all these requirements are consistent with cost-effective low-energy performance, and in the real detector some high-energy capacity may be compromised. 

More studies are required to draw more definite conclusions and to quantify the dependence on detector properties and optimization. 
\\

\noindent
{\bf \Large Acknowledgement}\\

This research was supported by the DFG cluster of excellence "Origin and Structure of the Universe" and
EU project LAGUNA. I thank all the LENA collaboration, particularly Mihcael Wurm, John Learned and other members of HanoHano and Silvia Pascoli for discussions, co-operation and advice.

\end{document}